\documentclass[letterpaper, 10 pt, conference]{ieeeconf}  

\IEEEoverridecommandlockouts                              
\usepackage[utf8]{inputenc}
\usepackage[T1]{fontenc}
\usepackage{subfigure}
\usepackage{amssymb,amsmath,bm}
\usepackage{textcomp}
\usepackage{amssymb,amsmath,epsfig,mdwlist,setspace}
\usepackage{array,multirow,color,booktabs,textcomp}
\usepackage{amsmath,graphicx}
\usepackage{CJKutf8}
\usepackage{amsfonts}
\usepackage{setspace}
\usepackage{etoolbox}

\title{\LARGE \bf
Behavior Score-Embedded Brain Encoder Network for Improved Classification of Alzheimer Disease Using Resting State fMRI
}

\author{Wan-Ting Hsieh$^{1,2,*}$, Jeremy Lefort-Besnard$^{1,2*}$, Hao-Chun Yang$^{1,2}$, Li-Wei Kuo$^{3}$, Chi-Chun Lee$^{1,2}$, \\ and the Alzheimer's Disease Neuroimaging Initiative$^{a}$
\thanks{$^{1}$Department of Electrical Engineering, National Tsing Hua University, Taiwan
        {\tt\small cclee@ee.nthu.edu.tw}}%
\thanks{$^{2}$MOST Joint Research Center for AI Technology and All Vista Healthcare, Taiwan}
\thanks{$^{3}$Institute of Biomedical Engineering and Nanomedicine, National Health Research Institutes, Taiwan}
\thanks{$^{a}$Data used in preparation of this article were obtained from the Alzheimer’s Disease Neuroimaging Initiative (ADNI) database (adni.loni.usc.edu). As such, the investigators within the ADNI contributed to the design and implementation of ADNI and/or provided data but did not participate in analysis or writing of this report.A complete listing of ADNI investigators can be found at: http://adni.loni.usc.edu/wp-content/uploads/how\_to\_apply/ADNI\_Acknowledgement\_List.pd}
\thanks{$^{*}$These authors contributed equally to this work}}%

\begin{document}

\maketitle
\thispagestyle{empty}
\pagestyle{empty}

\begin{abstract}

The ability to accurately detect onset of dementia is important in the treatment of the disease. Clinically, the diagnosis of Alzheimer Disease (AD) and Mild Cognitive Impairment (MCI) patients are based on an integrated assessment of psychological tests and brain imaging such as positron emission tomography (PET) and anatomical magnetic resonance imaging (MRI). In this work using two different datasets, we propose a behavior score-embedded encoder network ($BSEN$) that integrates regularly adminstrated psychological tests information into the encoding procedure of representing subject's resting-state fMRI data for automatic classification tasks. $BSEN$ is based on a 3D convolutional autoencoder structure with contrastive loss jointly optimized using behavior scores from Mini-Mental State Examination (MMSE) and Clinical Dementia Rating (CDR). Our proposed classification framework of using $BSEN$ achieved an overall recognition accuracy of 59.44\% (3-class classification: AD, MCI and Healthy Control), and we further extracted the most discriminative regions between healthy control (HC) and AD patients.

\end{abstract}
\vspace{-2mm}

\section{INTRODUCTION}\vspace{-1mm}

Dementia is a multifaceted mental disorder characterized by behavioral changes, cognitive deficits, and functional deteriorations. While Alzheimer Disease (AD) is the most common form of dementia occurred in elderly people, research into understanding Mild Cognitive Impairment (MCI) has gained interest since it is considered as an intermediate stage between normal aging and AD \cite{jessen2019we}. Due to the irreversible nature of dementia, early detection of AD at its pre-clinical stage is important for intervention. Numerous distinct psychological tests, such as Mini-Mental State Examination (MMSE) and Clinical Dementia Rating (CDR), have shown to be efficient to pre-screen subjects with major cognitive and memory impairment \cite{yagi2019identification}. The numerical scores obtained from the subject's self-report answers to these testings are considered as \textit{behavior indicators} reflecting neuro-psychological markers of potential impairment. These results are thus regularly adminstrated to assist in the decision of proceeding with the actual invasive clinical diagnosis \cite{american2013diagnostic}.\vspace{-0.5mm}

In the last decade, functional Magnetic Resonance Imaging (fMRI) has become a prevalent non-invasive method in the study of brain's functional activity. Resting-state functional activity is of particular interest since researchers have found that low-frequency blood oxygenation level dependent (BOLD) signal fluctuation during subject's resting condition shows a high synchronization pattern within motor cortices, visual cortices, language area and default mode network (DMN) \cite{zou2008improved}. Furthermore, resting-state fMRI (rs-fMRI) has been extensively employed in studies of brain's functional connectivity in AD and MCI subjects. For example, Wang et al.\ reported that the functional connectivity between right hippocampus and regions in DMN was disrupted in AD \cite{wang2006changes}, and they further discovered a reduced integrity in the thalamus-related cortical networks for MCI subjects \cite{wang2012changes}.\vspace{-0.5mm}

Aside from these studies, researcher have further shown that by using machine learning (ML) techniques, subjects of AD, MCI and HC (healthy control) can be automatically differentiated using their rs-fMRI data respectively. This provides a viable mean in dementia classification using non-invasive bio-imaging technique. For example, Khazaee et al.\ utilized the Granger causality measures with naïve Bayes classifier to identify AD, MCI and HC \cite{khazaee2017classification}; Luo et al.\ applied a 3D convolutional neural network (CNN) to extract the representation of brain to differentiate AD patients from HC \cite{luo2017automatic}. Clinically, the final diagnosis of dementia relies on comprehensive assessment including both psychological tests and brain imaging (MRI and computerized tomography), however, most of these previous automatic classification works using rs-fMRI do not integrate these behavior scores in their framework. In this work, our goal is to embed this important auxiliary information directly into a brain encoder network to generate rs-fMRI representation that would improve dementia classification from brain imaging.\vspace{-0.5mm}

Specifically, we propose a behavior score-embedded brain encoder network ($BSEN$) that can be used to encode rs-fMRI as embedding for dementia discriminative task. The $BSEN$ is based on a 3D CNN autoencoder (AE) architecture that is jointly optimized with contrastive loss \cite{qi2017contrastive} guided by the psychological testing scores. The encoder portion of the $BSEN$, once learned, can then be used to generate representation on patient's rs-fMRI data without asking the subject to work through these psychological testings. Our method achieved an improvement of almost 10\% over vanilla 3D-CNN AE in the diagnosis prediction. Our study was further replicated using another dataset and the obtained results that were virtually identical. Finally, in a follow-up analysis, we demonstrated that there exist distinctive ROIs being included in embedding between subjects of HC and patients using our $BSEN$ extractor.
\vspace{-1mm}

\section{RESEARCH METHODOLOGY}

\subsection{fMRI Data Collection and Preprocessing}

In the first set ($analysis$ $set$), raw rs-fMRI data consisting of 26 HC subjects, 23 MCI subjects and 21 AD subjects were collected. fMRI scanning was performed on 1.5 T scanner. A high-resolution T1-weighted 3D-SPGR anatomical scan was acquired for co-registration between structural and functional images (TR/TE = 3000/35 ms, voxel size = 3$\times$3$\times$3 ${mm}^{3}$, 43 slices, 120 repetitions, and flip angle =  ${90}^{\circ}$). Data used for replication of this study ($replication$ $set$) were obtained from the ADNI database (www.loni.ucla.edu/ADNI) including 29 HC subjects, 4 MCI subjects and 21 AD subjects (see Table 1 for demographics information of the 2 datasets). Note that the ADNI dataset has most often been used for its structural data, here, the resting-state fMRI scans were used. All image acquisitions have been carefully quality-controlled by experienced neuroimaging investigators and some of the subjects in the ADNI dataset were removed because of the low quality of images. fMRI scanning was performed on 3.0 T scanner (TR/TE = 3000/35 ms, voxel size: 3.3125$\times$3.3125$\times$3.3125 ${mm}^{3}$, 48 slices, 140 repetitions, and flip angle =  ${80}^{\circ}$). In both sets, The data were subjected to a standard resting-state preprocessing pipeline using SPM12 \cite{friston1995spatial} and DPARSF \cite{yan2010dparsf} including slicing timing and realignment. We used the images from ${20}^{th}$ time point to ${90}^{th}$ to prevent potential noisy data. The Institutional Review Board of National Health Research Institute Taiwan approved the study.

\begin{table}[t]
\caption{\it Demographic and clinical information}
\vspace{-0.01mm}
\label{tabel1}
\scalebox{0.56}{
\begin{tabular}{lcccl}
\toprule[1.5pt]
\textbf{Analysis set}              & HC(26)                 & MCI(23)                & AD(21)                 &  \\
Female / Male                       & 21 / 5                 & 17 / 6                 & 17 / 4                 &  \\
Age (mean$\pm$std) {[}min-max{]}       & 63.04$\pm$5.60 {[}55-72{]} & 70.78$\pm$8.34 {[}58-86{]} & 76.36$\pm$7.67 {[}68-88{]} &  \\
Education (mean$\pm$std) {[}min-max{]} & 10.54$\pm$4.42 {[}3-16{]}  & 6.83$\pm$4.30 {[}0-16{]}   & 4.45$\pm$4.72 {[}0-18{]}   &  \\
CDR (mean$\pm$std) {[}min-max{]}       & 0.03$\pm$0.14 {[}0-0.5{]}  & 0.24$\pm$0.33 {[}0-2{]}    & 0.89$\pm$0.45 {[}0.5-2{]}  &  \\
MMSE (mean$\pm$std) {[}min-max{]}      & 27.58$\pm$2.21 {[}23-30{]} & 24.70$\pm$3.83 {[}5-29{]}  & 14.48$\pm$6.46 {[}6-27{]}  &  \\ \hline
\textbf{Replication set}           & HC(29)                 & MCI(4)                & AD(21)                 &  \\
Female / Male                       & 11 / 18                 & 1 / 3                 & 12 / 9                 &  \\
Age (mean$\pm$std) {[}min-max{]}       & 76.02$\pm$7.54 {[}65-94{]} & 71.87$\pm$9.77 {[}63-88{]} & 74.0$\pm$7.41 {[}56-86{]} &  \\
Education                              &                          & no available information  &  \\
CDR (mean$\pm$std) {[}min-max{]}       & 0.12$\pm$0.25 {[}0-1{]}  & 0.5$\pm$0 {[}0.5-0.5{]}    & 1.02$\pm$0.44 {[}0.5-2{]}  &  \\
MMSE (mean$\pm$std) {[}min-max{]}      & 29.17$\pm$1.41 {[}24-30{]} & 28.25$\pm$1.48 {[}26-30{]}  & 21.09$\pm$3.24 {[}15-27{]}  &  \\
\bottomrule[1.5pt]
\end{tabular}}\vspace{-3mm}
\end{table}
\vspace{-2mm}

\subsection{Psychological Tests}\vspace{-1mm}
In this work, we use behavior scores generated from the two psychological testings, Clinical Dementia Rating (CDR) and Mini-Mental State Examination (MMSE), to guide the learning of our encoder network for rs-fMRI data.

\begin{itemize}
\item CDR: It rates the cognitive performance in six domains: memory, orientation, judgment and problem solving, community affairs, home and hobbies, and personal care. Each domain is rated independently from one to five indicating levels of impairment. In this work, we binarize the CDR into healthy cluster and probable dementia using CDR $=$ 0.5 as cut-off.
\item MMSE: It estimates the severity of cognitive impairment in seven categories: orientation to time, orientation to place, registration of three words, attention and calculation, recall of three words, language, and visual construction. The total score is 30 points. We binarize the MMSE into healthy and impaired group with the cut-off score of MMSE $=$ 27.
\end{itemize}

\vspace{-2mm}

\subsection{Behavior Score-Embedded Encoder Network ($BSEN$)}

\begin{figure}[t]
\raggedright
\center
\includegraphics[scale=0.4]{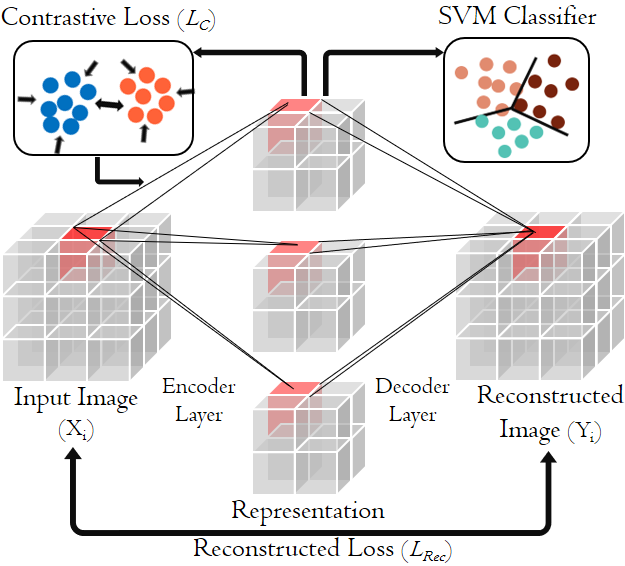}
\vspace*{-4 mm}
\caption{\it A schematic of the BSEN architecture in classifying HC, MCI and AD. The input of the network is BOLD signal of subjects per time.
\label{fig:scoreinfo}}\vspace{-8mm}

\end{figure}

Figure 1 depicts schematic of our proposed behavior score-embedded encoder network ($BSEN$). The $BSEN$ is a representation (feature) extractor with two stage optimization: first is an autoencoder (${model}_{AE}$), and second is contrastive loss model (${model}_{C}$). The details of each network learning are described below:
\begin{itemize}
\item The convolutional autoencoder (${model}_{AE}$): We employ an 3D-CNN autoencoder architecture to learn the encoder and decoder network. Given a collection of N training sample pairs \{${X}_{i}$, ${Y}_{i}$\}, where ${X}_{i}$ is a reconstructed brain image and ${Y}_{i}$ is the original brain image. We minimize the following Mean Squared Error (MSE):

   \vspace{-2mm}\begin{equation}
    \mathcal{L}_{Rec} = \frac{1}{N}\sum\limits_{i=1}^N {\left\|{X}_{i}-{Y}_{i}\right\|}_{2}^2
    \vspace{-1mm}\end{equation}

\item The contrastive loss model (${model}_{C}$): We embed the information of psychological tests as an additional loss term, $\mathcal{L}_{C}$, to modify the originally learned CNN AE to achieve a more discriminative hidden representation. The objective of the contrastive loss aims at enhancing intra-class compactness and inter-class dispersion. In this case, this loss acts as a regularizer forcing rs-fMRI representation with similar behavior scores (the two different group specified according to the cutoff score in section B) would be encoded as having more similar latent vectors than otherwise. This loss is embedded to the bottleneck layer of the CNN AE by explicitly centering the representation with respect to each behavior score center in the following form:

    \vspace{-2mm}\begin{equation}
    \mathcal{L}_{C} = \frac{1}{2}\sum\limits_{i=1}^m \frac{{\left\|{x}_{i}-{c}_{{e}_{i}}\right\|}_{2}^2}{({\sum_{j=1,j\neq e_i}^m\left\|{x}_{i}-{c}_{j}\right\|}_{2}^2)+\delta}
   \vspace{-1mm}\end{equation}

    where $m$ is the number of clusters (i.e., 2 for each psychological tests), ${x}_{i}$ is the data under condition $i$ and ${c}_{j}$ is the center of $j$ cluster (${c}_{{e}_{i}}$ denotes to the similar meaning of cluster ${e}_{i}$ ). $\delta$ is set as 1 preventing the denominator equals to 0.
\end{itemize}

Finally, the complete $BSEN$ is optimized using a total loss, $\mathcal{L}_{Total}$, with weighted hyperparameter $\alpha$ set as 0.5 in the following form:
\vspace{-2mm}\begin{equation}
\mathcal{L}_{total} = \mathcal{L}_{Rec}+\alpha\mathcal{L}_{C}
\vspace{-2mm}\end{equation}

\subsection{Dementia Classification and Fusion Technique}\vspace{-2mm}
We derived the rs-fMRI feature by extracting $BSEN$ output from the bottleneck layer for each subject as input to the multi-class linear-kernel support vector machine for HC, MCI and AD classification. In addition, we conducted decision score late fusion to perform ensemble on two $BSENs$ each learned from a different psychological tests.

\section{EXPERIMENTAL SETUP AND RESULTS}

We carried out three-class classification (HC, MCI and AD) task. The evaluation scheme was via 5-fold cross-validation. The accuracy was measured in unweighted average recall (UAR). All of the network learning and additional feature selection were carried out only in the training set.
\vspace{-2mm}

\subsection{Experimental Setup}\vspace{-1mm}
The $BSEN$ architecture is composed of three 3D convolutional layers with numbers of channels 32-16-8 for encoder and symmetrically for decoder. The kernel size of layers was set as $3\times3\times3$, while stride was set at 1$\times$1$\times$1 and zero padding were 1$\times$1$\times$1 to ensure the consistent dimensions. Each convolutional layer was followed with batch normalization and maxpooling with size 2$\times$2$\times$2. At the last layer of decoder, the activation function ReLU was used. The total loss function is composed of contrastive loss on 5120-dimensional latent bottleneck representation and mean squared error for reconstruction task. The batch size and epoch were set to 32 and 30 respectively. $BSEN$ is optimized using Adam with learning rate = 0.0005 (contrastive learning) and 0.0001 (autoencoder learning).  We paded the raw data, $61\times73\times61$ for the $analysis$ $set$, and $60\times72\times60$ for the $replication$ $set$, to respectively $64\times80\times64$ and $64\times72\times64$ with zeros in order to better perform maxpooling and upsampling. Once the $BSEN$ was trained, the subject's averaged brain image was fed into the $BSEN$ encoder, and the bottleneck layer was extracted (8 channels). The feature representation of subjects was obtained by mean pooling the bottleneck layer across channels.

We compared our framework with the following methods to derive rs-fMRI feature representation:
\begin{itemize*}

\item \textbf{ICA}: Perform independent component analysis on BOLD signal.
\item \textbf{PCA}: Perform principal component analysis on BOLD signal.
\item \textbf{CAE}: Perform 3D convolutional autoencoder without contrastive loss embedding.
\end{itemize*}
Our proposed $BSEN$ was learned with two different psychological tests, ${B}_{CDR}SEN$, ${B}_{MMSE}SEN$. These features extracted from the above-mentioned models were then fed into the HC, MCI and AD classification procedures. ${BSEN}_{Fusion}$ indicates the final late fusion from the two $BSEN$ models. Please, note that all analysis scripts of the present study are readily accessible to the reader online  (https://biicgitlab.ee.nthu.edu.tw/JeremyLB/bsen).
\vspace{-1mm}
\subsection{Experimental Results and Discussion}\vspace{-1mm}

\begin{table}[t]
\vspace{-1mm}
\caption{\it 3-class classification results of our proposed BSEN model and other representation techniques. The accuracy is measured in UAR (\%).}
\vspace{-0.1mm}
\scalebox{0.65}{
\raggedright
\begin{tabular}{lccccccc}
\toprule[1.5pt]
    & $ICA$   & $PCA$   & $CAE$   & ${B}_{CDR}SEN$ & ${B}_{MMSE}SEN$ & ${BSEN}_{Fusion}$ \\ \hline
HC  & 46.15 & 42.31 & 61.54 & 57.69    & 57.69     &    61.54         \\
MCI & 39.13 & 65.22 & 56.52 & 56.52    & 69.57     &    73.91      \\
AD  & 42.85 & 38.09 & 33.33 & 52.38    & 38.09     &    42.86        \\ \hline
UAR & 42.71 & 48.54 & 50.46 & 55.53    & 55.11     & \textbf{59.44} \\ \hline
Replication & 38.62 & 44.44 & 48.23 & 52.52    & 52.13     & \textbf{56.32} \\
\bottomrule[1.5pt]
\end{tabular}}
\end{table}

\begin{table}[]
\caption{\it Significantly different ROIs discovered from two-sided Students t-tests between HC and AD.}
\scalebox{0.7}{
\centering
\begin{tabular}{lll}
\hline
\multicolumn{3}{l}{\textbf{ROIs found in CAE and $BSEN$}}                                         \\ \hline
\textit{ROIs}                                            & \textit{T-value}  & \textit{P-value} \\
Anterior cingulate cortex (R/L)                          & 2.183/2.25        & 0.034/0.029      \\
Insula (L)                                               & 2.055             & 0.046            \\
Middle frontal gyrus (L)                                 & 2.106             & 0.041            \\
Inferior frontal gyrus, triangular (L)                   & 2.447             & 0.018            \\
Inferior frontal gyrus, orbital (L)                      & 2.202             & 0.033            \\
Rolandic operculum (L)                                   & 2.113             & 0.04             \\
Superior frontal gyrus, medial (L)                       & 2.216             & 0.032            \\
Superior temporal gyrus, temporal pole (L)               & 2.239             & 0.03             \\ \hline
\multicolumn{3}{l}{\textbf{ROIs found in $BSEN$ only}}                                            \\ \hline
\multirow{2}{*}{Precentral gyrus (R/L)}                  & CDR: 2.806/2.341  & 0.007/0.024      \\
                                                         & MMSE: 2.956/2.761 & 0.005/0.008      \\
\multirow{2}{*}{Inferior frontal gyrus, opercular (R/L)} & CDR: 2.053/X      & 0.046/X          \\
                                                         & MMSE: 2.298/2.272 & 0.026/0.028      \\
\multirow{2}{*}{Supplementary motor area (R/L)}          & CDR: 2.364/2.140  & 0.023/0.039      \\
                                                         & MMSE: 2.143/X     & 0.039/X          \\ \hline
\multicolumn{3}{l}{\textbf{ROIs found in ${B}_{CDR}SEN$ only}}                                       \\ \hline
Paracentral lobule (R)                                   & 2.134             & 0.039            \\
Supramarginal gyrus (L)                                  & 2.116             & 0.04             \\
Inferior parietal gyrus (L)                              & 2.114             & 0.04             \\
Angular gyrus (L)                                        & 2.109             & 0.041            \\ \hline
\multicolumn{3}{l}{\textbf{ROIs found in ${B}_{MMSE}SEN$ only}}                                    \\ \hline
Superior temporal gyrus, temporal pole (R)               & 2.604             & 0.013            \\
Middle temporal gyrus (L)                                & 2.051             & 0.046           
\end{tabular}}\vspace{-7mm}
\end{table}

Table 2 summarizes our complete experimental results. The proposed $BSEN$ obtained the best accuracy compared to all other baseline systems. Especially, when using representation learned with CDR score, which received 55.53 \% UAR (almost 20\% above chance level). We further performed late fusion by weighted summing of the predicted probabilities from the two behavior score-embedded representations, which improved UAR to 59.44\%. When comparing with all of the baseline systems, performances achieved using our proposed $BSEN$ were better than all other benchmarks demonstrating that the behavior score embedding via contrastive joint optimization can indeed help improve discriminative power in dementia classification using rs-fMRI. Finally, in a replication analysis, virtually identical results were obtained.

In summary, using the convolutional autoencoder approach (CAE), which can been seen as a non-linear version of PCA, improved slightly the recognition results (by 1.92\%). The major boost in the accuracy comes from integrating psychological testing information, which increased the overall performance up to almost 10\% when comparing with CAE.

\vspace{-1mm}
\section{Discriminative regions analysis}

To further understand how neuropsychological tests were involved during the embedding learning process, we reconstructed the brain rs-fMRI representations of our cohort for each group (AD, MCI, HC) using the decoder portion of the $BSEN$. We assessed statistical significance based on (family
wise error, multiple-comparison corrected) p-values on the reconstructed brain images to analyze whether the active ROIs (defined by the  automated anatomical labeling (AAL)) would demonstrate significantly different activation between HC and AD. Table 3 summarizes the significant ROIs when comparing HC with AD.

Part of the salience network, critical for guidance of thought and behaviors, was found to be aberrant in patients with AD compared with HC and MCI patients as in a previous work \cite{agosta2012resting}. Besides, rs-fMRI recording show abnormal activation in several brain areas in AD, such as middle frontal gyrus, inferior frontal gyrus (both triangular and orbital), rolandic operculum, superior frontal gyrus (medial) and superior temporal gyrus (temporal pole) \cite{wang2007altered}. In this work, these ROIs were also identified to be the dominant ROIs in the derived discriminative representation extracted from both CAE and $BSEN$. In addition, bilateral precentral gyrus, inferior frontal gyrus (opercular), and supplementary motor area, that were additionally identified in the two behavior score-embedded model (${B}_{CDR}SEN$, ${B}_{MMSE}SEN$), are likely to be the major contributing factor that $BSEN$ would outperform CAE.

It is worth noting that there was a difference between the two psychological test-enhanced representations. For example, representations of right paracentral lobe and left angular gyrus demonstrated considerable differences between HC and AD when embedded with CDR scores while the left middle temporal and the right superior temporal gyras were used as discriminant when embedded MMSE scores. Each of these psychological tests, while mostly common, may still be related to distinct functional activation in subject's brain during resting state; hence, fusing all of these representations helped achieving the improvement in our results.
\vspace{-1mm}

\section{CONCLUSIONS}\vspace{-1mm}
In this work, we introduced a novel feature embedding framework of brain imaging for classifying HC, MCI and AD. Specifically, we propose a $BSEN$ to encode the subject’s rs-fMRI through jointly optimizing behavior scores using contrastive loss embedding. We achieve a promising 59.44\% accuracy in three-way classification. Our future work is to investigate the differences observed when learning rs-fMRI representations with the two psychological tests to further tease apart the underlying neuro-cognitive mechanistic differences between AD, MCI and HC.

\bibliographystyle{IEEEtran}
\bibliography{IEEEabrv,refs}

\begin{thebibliography}{10}
\providecommand{\url}[1]{#1}
\csname url@samestyle\endcsname
\providecommand{\newblock}{\relax}
\providecommand{\bibinfo}[2]{#2}
\providecommand{\BIBentrySTDinterwordspacing}{\spaceskip=0pt\relax}
\providecommand{\BIBentryALTinterwordstretchfactor}{4}
\providecommand{\BIBentryALTinterwordspacing}{\spaceskip=\fontdimen2\font plus
\BIBentryALTinterwordstretchfactor\fontdimen3\font minus
  \fontdimen4\font\relax}
\providecommand{\BIBforeignlanguage}[2]{{%
\expandafter\ifx\csname l@#1\endcsname\relax
\typeout{** WARNING: IEEEtran.bst: No hyphenation pattern has been}%
\typeout{** loaded for the language `#1'. Using the pattern for}%
\typeout{** the default language instead.}%
\else
\language=\csname l@#1\endcsname
\fi
#2}}
\providecommand{\BIBdecl}{\relax}
\BIBdecl

\bibitem{jessen2019we}
F.~Jessen, ``What are we trying to prevent in alzheimer disease?''
  \emph{DIALOGUES IN CLINICAL NEUROSCIENCE}, vol.~21, no.~1, pp. 27--34, 2019.

\bibitem{yagi2019identification}
T.~Yagi, M.~Kanekiyo, J.~Ito, R.~Ihara, K.~Suzuki, A.~Iwata, T.~Iwatsubo,
  K.~Aoshima, A.~D.~N. Initiative, J.~A. D.~N. Initiative \emph{et~al.},
  ``Identification of prognostic factors to predict cognitive decline of
  patients with early alzheimer's disease in the japanese alzheimer's disease
  neuroimaging initiative study,'' \emph{Alzheimer's \& Dementia: Translational
  Research \& Clinical Interventions}, vol.~5, pp. 364--373, 2019.

\bibitem{american2013diagnostic}
A.~P. Association \emph{et~al.}, \emph{Diagnostic and statistical manual of
  mental disorders (DSM-5{\textregistered})}.\hskip 1em plus 0.5em minus
  0.4em\relax American Psychiatric Pub, 2013.

\bibitem{zou2008improved}
Q.-H. Zou, C.-Z. Zhu, Y.~Yang, X.-N. Zuo, X.-Y. Long, Q.-J. Cao, Y.-F. Wang,
  and Y.-F. Zang, ``An improved approach to detection of amplitude of
  low-frequency fluctuation (alff) for resting-state fmri: fractional alff,''
  \emph{Journal of neuroscience methods}, vol. 172, no.~1, pp. 137--141, 2008.

\bibitem{wang2006changes}
L.~Wang, Y.~Zang, Y.~He, M.~Liang, X.~Zhang, L.~Tian, T.~Wu, T.~Jiang, and
  K.~Li, ``Changes in hippocampal connectivity in the early stages of
  alzheimer's disease: evidence from resting state fmri,'' \emph{Neuroimage},
  vol.~31, no.~2, pp. 496--504, 2006.

\bibitem{wang2012changes}
Z.~Wang, X.~Jia, P.~Liang, Z.~Qi, Y.~Yang, W.~Zhou, and K.~Li, ``Changes in
  thalamus connectivity in mild cognitive impairment: evidence from resting
  state fmri,'' \emph{European journal of radiology}, vol.~81, no.~2, pp.
  277--285, 2012.

\bibitem{khazaee2017classification}
A.~Khazaee, A.~Ebrahimzadeh, A.~Babajani-Feremi, A.~D.~N. Initiative
  \emph{et~al.}, ``Classification of patients with mci and ad from healthy
  controls using directed graph measures of resting-state fmri,''
  \emph{Behavioural brain research}, vol. 322, pp. 339--350, 2017.

\bibitem{luo2017automatic}
S.~Luo, X.~Li, and J.~Li, ``Automatic alzheimer’s disease recognition from
  mri data using deep learning method,'' \emph{Journal of Applied Mathematics
  and Physics}, vol.~5, no.~09, p. 1892, 2017.

\bibitem{qi2017contrastive}
C.~Qi and F.~Su, ``Contrastive-center loss for deep neural networks,''
  \emph{arXiv preprint arXiv:1707.07391}, 2017.

\bibitem{friston1995spatial}
K.~J. Friston, J.~Ashburner, C.~D. Frith, J.-B. Poline, J.~D. Heather, and
  R.~S. Frackowiak, ``Spatial registration and normalization of images,''
  \emph{Human brain mapping}, vol.~3, no.~3, pp. 165--189, 1995.

\bibitem{yan2010dparsf}
C.~Yan and Y.~Zang, ``Dparsf: a matlab toolbox for" pipeline" data analysis of
  resting-state fmri,'' \emph{Frontiers in systems neuroscience}, vol.~4,
  p.~13, 2010.

\bibitem{agosta2012resting}
F.~Agosta, M.~Pievani, C.~Geroldi, M.~Copetti, G.~B. Frisoni, and M.~Filippi,
  ``Resting state fmri in alzheimer's disease: beyond the default mode
  network,'' \emph{Neurobiology of aging}, vol.~33, no.~8, pp. 1564--1578,
  2012.

\bibitem{wang2007altered}
K.~Wang, M.~Liang, L.~Wang, L.~Tian, X.~Zhang, K.~Li, and T.~Jiang, ``Altered
  functional connectivity in early alzheimer's disease: A resting-state fmri
  study,'' \emph{Human brain mapping}, vol.~28, no.~10, pp. 967--978, 2007.

\end{thebibliography}
\end{document}